# Schrödinger potentials solvable in terms of the general Heun functions

A.M. Ishkhanyan[1,2]

[1]Institute of Physics and Technology, National Research Tomsk Polytechnic University, Tomsk 634050, Russia
[2]Institute for Physical Research, NAS of Armenia, Ashtarak 0203, Armenia

We show that there exist 35 choices for the coordinate transformation each leading to a potential for which the stationary Schrödinger equation is exactly solvable in terms of the general Heun functions. Because of the symmetry of the Heun equation with respect to the transposition of its singularities only eleven of these potentials are independent. Four of these independent potentials are always explicitly written in terms of elementary functions, one potential is given through the Jacobi elliptic sn-function, and the others are in general defined parametrically. Nine of the independent potentials possess exactly or conditionally integrable hypergeometric sub-potentials for which each of the fundamental solutions of the Schrödinger equation is written through a single hypergeometric function. Many of the potentials possess sub-potentials for which the general solution is written through fundamental solutions each of which is a linear combination of two or more Gauss hypergeometric functions. We present an example of such a potential which is a conditionally integrable generalization of the third exactly solvable Gauss hypergeometric potential.

## 1. Introduction

The Schrödinger equation describes many quantum phenomena and has an extremely large range of applications in contemporary physics ranging from particle, atomic, molecular and optical physics to general gravity, astrophysics and cosmology. For problems with time-independent Hamiltonians useful particular solutions can be constructed via separation of variables, that is, by reduction of the problem to the solution of the stationary version of the Schrödinger equation. This approach was widely applied in the past using the functions of the hypergeometric class. Some of the field potentials, such as the Coulomb [1], harmonic oscillator [1], centrifugal [1], Kratzer [2], Morse [3], Rosen-Morse [4], Manning-Rosen [5], Hulthén [6], Woods-Saxon [7], Eckart [8], Scarf [9], Pöschl-Teller [10] ones, that allow the solution of the Schrödinger equation in terms of these functions became standard tools for demonstration and exploration of the handful of the basic quantum paradigms.

However, though the seemingly large set of reported potentials, the list of the independent classical hypergeometric potentials, both ordinary and confluent, is in fact rather short. Indeed, the classical ordinary hypergeometric potentials all are particular specifications of only two independent ones, the Eckart [8] and Pöschl-Teller [10] potentials, and the list of independent classical confluent hypergeometric potentials includes just three names, the Coulomb (plus centrifugal) and the harmonic oscillator (plus inverse square) potentials discussed by Schrödinger in establishing the new wave mechanics [1] and the Morse potential that was reported soon after the Schrödinger equation appeared [3].

The need to go beyond the hypergeometric class of functions in order to derive new solvable potentials has been noticed soon after the classical hypergeometric potentials have been explored. First Morse and Stückelberg [11] and then Manning [12] have proposed important potentials that do not fall into the hypergeometric class. Further, classifying and analyzing the available results, Manning has discussed the reduction of the Schrödinger equation to a rather large class of target equations having rational coefficients and indicated some of the ones that are the immediate next to the hypergeometric equations [13]. A main message by Morse, Stückelberg and Manning is that one should appeal to equations the series solutions to which are governed by three or more term recurrence relations, instead of the hypergeometric two-term ones, for the coefficients of the successive terms of expansions.

Many authors have afterwards explored this possibility. In 1969 Lemieux and Bose gave a first systematic treatment using certain equations the solutions of which are governed by three-term recurrence relations for coefficients of power-series expansions [14]. These equations are known to compose the Heun class of equations [15-17] that present diverse generalizations of the equations of the hypergeometric class as well as several other known equations including the Mathieu, spheroidal and Coulomb spheroidal equations [18].

Though the paper by Lemieux and Bose was not noticed much, however, it contains a rather advanced analysis of a large class of potentials solvable in terms of the Heun functions. The work was preceded by a careful treatment of the hypergeometric potentials by Bose [19,20] who applied an advanced approach based on the *invariant* of the Liouville *normal form* of the target equation to which the Schrödinger equation is reduced and the *Schwartzian derivative* of the applied coordinate transformation. The Bose approach was afterwards employed by Natanzon to establish the general seven-parametric form of the potentials, constructed by energy-independent coordinate transformations, solvable in terms of the ordinary and confluent hypergeometric functions [21,22].

It has recently been shown, however, that under rather general conditions this general Natanzon potential is dropped into a restricted discrete set of sub-potentials involving a lesser number of parameters [23]. Apart from this general result, the treatment of the discretization of the Natanzon potentials introduced some new features in the technique for searching for the exactly solvable potentials. These are rather productive features. For instance, using this technique three new potentials exactly solvable in terms of the confluent hypergeometric functions (the inverse square root potential [24], the Lambert-W step potential [25] and the Lambert-W singular potential [26]) as well as the third exactly solvable Gauss hypergeometric potential [27] have been reported. Notably, the approach also allows one to derive conditionally integrable potentials that may have useful applications in structurally analogous problems where weaker restrictions can be imposed on the applicable potentials, e.g., in quantum optics [28-30]. Representative examples of this type of potentials are presented in [25,31,32] (see also [28]).

Having in the mind these new results and the developed approach, in the present paper we discuss the reduction of the Schrödinger equation to the general Heun equation which is a natural generalization of the Gauss hypergeometric equation. This is the most general ordinary linear second-order Fuchsian differential equation having four regular singular points. The special functions emerging from this equation and its four confluent forms are widely faced in current applied and fundamental physics and mathematics research (see, e.g., [15-17] and references therein).

If the reduction of the Schrödinger equation to the general Heun equation is supposed to be through an energy-independent coordinate transformation, one arrives at a general fourteen-parametric potential [33] (which is reduced to a twelve-parametric one if one applies the *canonical* form of the Heun equation with two of the singularities located at $z=0$ and $z=1$ [15-16]). We show that if one additionally requires the potential to be proportional to an energy-independent parameter and have a shape that is independent of that parameter, then there exist only thirty-five permissible forms for the coordinate transformation. Each of these cases leads to a ten-parametric potential (or eight-parametric if two of the singularities are fixed as $z=0,1$). Because of the symmetry of the general Heun equation with respect to the transposition of its singularities the number of independent potentials is reduced to eleven.

In general, the potentials are defined parametrically as pairs of functions $x(z), V(z)$; however, in several cases the coordinate transformation $x(z)$ is inverted thus producing explicitly written potentials given as $V = V(z(x))$ through a function $z = z(x)$. For five potentials the coordinate transformations are inverted for all involved parameters. Among these, in four cases the transformations are given in terms of elementary functions and in the fifth case the transformation is written through the Jacobi elliptic sn-function. It should be mentioned that though for the remaining six cases the coordinate transformation is generally not inverted, however there exist several particular cases with specific sets of parameters for which the transformation is nevertheless inverted.

The four independent potentials for which the inverted transformations are always written in terms of elementary functions present distinct generalizations of the classical five-parametric exactly integrable hypergeometric potentials, either the one by Eckart [8] or that by Pöschl and Teller [10]. These Heun potentials have been indicated by Lemieux and Bose [14]. We here indicate three other Heun potentials which also possess exactly solvable such classical hypergeometric sub-potentials. The most of the independent potentials possess *conditionally* exactly solvable hypergeometric sub-potentials. We list several examples of such potentials. It is worth mentioning that in many cases the general solution of the Schrödinger equation for these potentials is written through fundamental solutions each of which is given as a linear combination (generally, with non-constant coefficients) of two or more hypergeometric functions. To the best of our knowledge, the most of these potentials have not been discussed before. We present an example of a potential for which each of the fundamental solutions involves two Gauss hypergeometric functions. This is a conditionally integrable generalization of the third exactly solvable hypergeometric potential [27].

## 2. Discretization of the potentials

The one-dimensional stationary Schrödinger equation for a particle of mass $m$ and energy $E$ in a potential $V(x)$ is written as

$$\frac{d^2\psi}{dx^2} + \frac{2m}{\hbar^2}(E - V(x))\psi = 0. \tag{1}$$

Applying a transformation $z = z(x)$, we rewrite equation (1) for a new argument $z$:

$$\psi_{zz} + \frac{\rho_z}{\rho}\psi_z + \frac{2m}{\hbar^2}\frac{E - V(z)}{\rho^2}\psi = 0, \tag{2}$$

where (and hereafter) the lowercase Latin subscript denotes differentiation and $\rho = dz/dx$. The transformation of the dependent variable $\psi = \varphi(z)u(z)$ then reduces the problem to the following equation for a new dependent variable $u(z)$:

$$u_{zz} + \left(2\frac{\varphi_z}{\varphi} + \frac{\rho_z}{\rho}\right)u_z + \left(\frac{\varphi_{zz}}{\varphi} + \frac{\rho_z}{\rho}\frac{\varphi_z}{\varphi} + \frac{2m}{\hbar^2}\frac{E - V(z)}{\rho^2}\right)u = 0. \tag{3}$$

This equation becomes a target equation given as

$$u_{zz} + f(z)u_z + g(z)u = 0 \tag{4}$$

if
$$2\frac{\varphi_z}{\varphi}+\frac{\rho_z}{\rho}=f(z), \quad \frac{\varphi_{zz}}{\varphi}+\frac{\rho_z}{\rho}\frac{\varphi_z}{\varphi}+\frac{2m}{\hbar^2}\frac{E-V(z)}{\rho^2}=g(z). \quad (5)$$

Determining $\varphi(z)$ from the first of these two equations:

$$\varphi(z)=\rho^{-1/2}\exp\left(\frac{1}{2}\int f(z)dz\right), \quad (6)$$

and substituting this into the second one, we get

$$g-\frac{f_z}{2}-\frac{f^2}{4}=-\frac{1}{2}\left(\frac{\rho_z}{\rho}\right)_z-\frac{1}{4}\left(\frac{\rho_z}{\rho}\right)^2+\frac{2m}{\hbar^2}\frac{E-V(z)}{\rho^2}. \quad (7)$$

This equation is rewritten as

$$z'(x)I(z)+\frac{1}{2}\{z,x\}=\frac{2m}{\hbar^2}\frac{E-V(z)}{\rho^2}, \quad (8)$$

where $I(z)$ is the *invariant* of equation (4) and $\{z,x\}$ is the *Schwartzian derivative* of $z(x)$:

$$I(z)\equiv g-\frac{f_z}{2}-\frac{f^2}{4}, \quad \{z,x\}\equiv\left(\frac{z''(x)}{z'(x)}\right)'-\frac{1}{2}\left(\frac{z''(x)}{z'(x)}\right)^2. \quad (9)$$

Equation (8) is a productive equation that has been applied by many authors starting from the work by Bose [19]. Though equivalent, however, for our purposes it is convenient to directly apply equation (7). This is because in order to construct potentials that are proportional to an energy-independent parameter, according to the basic statements of [23], the first two terms in equation (7) involving the logarithmic derivative $\rho_z/\rho$ should be treated independently. This is a key step that allows one to identify the permissible coordinate transformations that lead to particular exactly solvable potentials that are proportional to an energy-independent parameter and have a shape that is independent of that parameter. It has been shown that the set of such permissible coordinate transformations is rather restricted, includes not so many possibilities [23-30].

To understand this, we first note that if the potential $V(x)$ is supposed to be energy independent and is constructed via an energy-independent coordinate transformation $z(x)$, then the energy term in equation (7), that is the function $1/\rho^2$, should independently match the left-hand side terms composing the invariant $I(z)$. This is immediately seen if one takes the first derivative of equation (7) with respect to $E$. Then, if the potential term is additionally supposed to be proportional to an energy-independent parameter $\mu$ and have a shape that does not depend on that parameter, it is shown that the coordinate transformation and thus the function $\rho(z)$ should necessarily be $\mu$-independent [23]. As a direct consequence, by taking the successive limits $E, \mu \to 0$, one arrives at the conclusion that the combination of the two terms involving $\rho_z / \rho$ should independently match the left-hand side invariant. It is then immediately seen that the logarithmic derivative $\rho_z / \rho$ cannot have poles other than the ones of the invariant $I(z)$ so that the coordinate transformation is necessarily of the Manning form [13]: $\rho(z) = \Pi_i (z - z_i)^{A_i}$ with $z_i$ being the finite singularities of the target equation (4) and exponents $A_i$ all being integers or half-integers.

For the general Heun equation:

$$u_{zz} + \left( \frac{\gamma}{z - a_1} + \frac{\delta}{z - a_2} + \frac{\varepsilon}{z - a_3} \right) u_z + \frac{\alpha \beta z - q}{(z - a_1)(z - a_2)(z - a_3)} u = 0, \qquad (10)$$

the invariant $I(z)$ is a fourth-degree polynomial in $z$ divided by $(z - a_1)^2 (z - a_2)^2 (z - a_3)^2$. Hence, taking into account that $E / \rho^2$ and $V(z) / \rho^2$ behave like $I(z)$ one has

$$\rho(z) \equiv z'(x) = \frac{\pm (z - a_1)(z - a_2)(z - a_3)}{\sqrt{r_0 + r_1 z + r_2 z^2 + r_3 z^3 + r_4 z^4}} \qquad (11)$$

and

$$\frac{2m}{\hbar^2} V(z) = \frac{v_0 + v_1 z + v_2 z^2 + v_3 z^3 + v_4 z^4}{r_0 + r_1 z + r_2 z^2 + r_3 z^3 + r_4 z^4} - \frac{1}{2} \{z, x\}. \qquad (12)$$

With arbitrary parameters $a_{1,2,3}$, $r_{0,1,2,3,4}$, $v_{0,1,2,3,4}$ and the integration constant of equation (11), this is a fourteen-parametric family of potentials. By scaling and shifting transformation $z \to z_0 + sz$, without loss of generality, one always may achieve $a_1 = 0$ and $a_2 = 1$, so that the number of the free parameters can be decreased to twelve. Furthermore, in accordance with the aforesaid, if one considers the case when the potential term in the Schrödinger equation is proportional to an energy-independent parameter and supposes that the potential shape is independent of both energy and that parameter, this general potential is dropped into a restricted set of sub-potentials involving considerably less independent parameters [23] (eight parameters if the canonical specification of the singularities of the general Heun equation is considered). We now present these potentials.

## 3. General Heun potentials

Thus, consider the case when $2mV(x) / \hbar^2 = \mu S(x)$ with $\mu \neq \mu(E)$, $S \neq S(\mu, E)$.

In accordance with the outlined treatment, the permissible $E$- and $\mu$-independent coordinate transformation for reduction of the Schrödinger equation to the general Heun equation is given through the equation

$$\rho = (z-a_1)^{m_1}(z-a_2)^{m_2}(z-a_3)^{m_3}/\sigma, \qquad (13)$$

where $m_{1,2,3}$ are integers or half-integers and $\sigma$ is an arbitrary scaling constant. To match the energy term $E/\rho^2$ with the left-hand side in equation (7), the ratio $z^2(z-1)^2(z-a)^2/\rho^2$ should be a polynomial in $z$ of at most fourth degree:

$$\sigma^2(z-a_1)^{2-2m_1}(z-a_2)^{2-2m_2}(z-a_3)^{2-2m_3} = r_0 + r_1 z + r_2 z^2 + r_3 z^3 + r_4 z^4 \stackrel{def}{=} r(z). \qquad (14)$$

This is achieved if $-1 \le m_{1,2,3} \le 1$ and $1 \le m_1 + m_2 + m_3 \le 3$. As a result, we get 35 possible triads $(m_1, m_2, m_3)$ shown in Fig. 1 by spheres in the 3D space of parameters $m_{1,2,3}$.

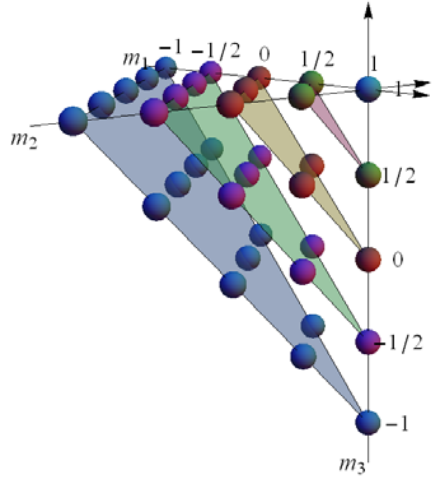

Fig. 1. 35 permissible triads $(m_1, m_2, m_3)$.

With these triads, matching the potential term $V(z)/\rho^2$ with the invariant in equation (7), we get 35 fourteen-parametric potentials which satisfy the equation

$$V(z)(r_0 + r_1 z + r_2 z^2 + r_3 z^3 + r_4 z^4) = v_0 + v_1 z + v_2 z^2 + v_3 z^3 + v_4 z^4 \stackrel{def}{=} v(z) \qquad (15)$$

with arbitrary $v_{0,1,2,3,4}$. We note that, because of the symmetry of the general Heun equation with respect to the transpositions of its singularities the number of independent potentials is eleven. These potentials are conveniently written as presented in Table 1. In five cases the coordinate transformation $x \to z$ is inverted so that in these cases the potentials are explicitly written as $V = V(z(x))$ through a function $z = z(x)$ shown in the fourth column of the table.

| N | $m_{1,2,3}$ | Potential $V(z)$ | Coordinate transformation $x(z)$ or $z(x)$ | |
|---|---|---|---|---|
| 1 | $1, 1, 1$ | $V_0 + V_1 z + V_2 z^2 + V_3 z^3 + V_4 z^4$ | $\dfrac{x - x_0}{\sigma} = \dfrac{(a_2 - a_3)\ln(z - a_1) + (a_3 - a_1)\ln(z - a_2) + (a_1 - a_2)\ln(z - a_3)}{(a_1 - a_2)(a_1 - a_3)(a_2 - a_3)}$ | $_2F_1(x^2)$ [8] |
| 2 | $1, 1, \tfrac{1}{2}$ | $\dfrac{V_0 + V_1 z + V_2 z^2 + V_3 z^3 + V_4 z^4}{z - a_3}$ | $\dfrac{x - x_0}{\sigma} = \dfrac{2}{a_1 - a_2}\left(\dfrac{1}{\sqrt{a_2 - a_3}}\tanh^{-1}\left(\sqrt{\dfrac{z - a_3}{a_2 - a_3}}\right) - \dfrac{1}{\sqrt{a_1 - a_3}}\tanh^{-1}\left(\sqrt{\dfrac{z - a_3}{a_1 - a_3}}\right)\right)$ | -- |
| 3 | $1, 1, 0$ | $\dfrac{V_0 + V_1 z + V_2 z^2 + V_3 z^3 + V_4 z^4}{(z - a_3)^2}$ | $z = \dfrac{a_1 - a_2 e^{(a_1 - a_2)(x - x_0)/\sigma}}{1 - e^{(a_1 - a_2)(x - x_0)/\sigma}}$, LB [14] | $_2F_1$ [8] |
| 4 | $1, 1, -\tfrac{1}{2}$ | $\dfrac{V_0 + V_1 z + V_2 z^2 + V_3 z^3 + V_4 z^4}{(z - a_3)^3}$ | $\dfrac{x - x_0}{\sigma} = \dfrac{2\sqrt{a_2 - a_3}}{a_1 - a_2}\left(\tanh^{-1}\left(\dfrac{\sqrt{z - a_3}}{\sqrt{a_2 - a_3}}\right) - \dfrac{\sqrt{a_1 - a_3}}{\sqrt{a_2 - a_3}}\tanh^{-1}\left(\dfrac{\sqrt{z - a_3}}{\sqrt{a_1 - a_3}}\right)\right)$ | CIP |
| 5 | $1, 1, -1$ | $\dfrac{V_0 + V_1 z + V_2 z^2 + V_3 z^3 + V_4 z^4}{(z - a_3)^4}$ | $\dfrac{x - x_0}{\sigma} = \dfrac{a_1 - a_3}{a_1 - a_2}\ln(z - a_1) + \dfrac{a_3 - a_2}{a_1 - a_2}\ln(z - a_2)$ | $_2F_1(x^2)$ [8] CIP |
| 6 | $1, \tfrac{1}{2}, \tfrac{1}{2}$ | $\dfrac{V_0 + V_1 z + V_2 z^2 + V_3 z^3 + V_4 z^4}{(z - a_2)(z - a_3)}$ | $\dfrac{x - x_0}{\sigma} = \dfrac{1}{\sqrt{a_1 - a_2}\sqrt{a_1 - a_3}}\left[\ln(a_1 - z) - \ln\left(\left(\sqrt{a_1 - a_2}\sqrt{z - a_2} + \sqrt{a_1 - a_3}\sqrt{z - a_3}\right)^2 - (a_2 - a_3)^2\right)\right]$ | $_2F_1(x^2)$ [10] |
| 7 | $1, \tfrac{1}{2}, 0$ | $\dfrac{V_0 + V_1 z + V_2 z^2 + V_3 z^3 + V_4 z^4}{(z - a_2)(z - a_3)^2}$ | $z = a_2 + (a_1 - a_2)\tanh\left(\dfrac{\sqrt{a_1 - a_2}(x - x_0)}{2\sigma}\right)^2$, LB [14] | $_2F_1$ [10] |
| 8 | $1, \tfrac{1}{2}, -\tfrac{1}{2}$ | $\dfrac{V_0 + V_1 z + V_2 z^2 + V_3 z^3 + V_4 z^4}{(z - a_2)(z - a_3)^3}$ | $\dfrac{x - x_0}{\sigma} = \ln\left(\left(\sqrt{z - a_2} + \sqrt{z - a_1}\right)^2\right) + \dfrac{\sqrt{a_1 - a_3}}{\sqrt{a_1 - a_2}}\left[\ln(a_1 - z) - \ln\left(\left(\sqrt{a_1 - a_2}\sqrt{z - a_2} + \sqrt{a_1 - a_3}\sqrt{z - a_3}\right)^2 - (a_2 - a_3)^2\right)\right]$ | CIP |
| 9 | $1, 0, 0$ | $\dfrac{V_0 + V_1 z + V_2 z^2 + V_3 z^3 + V_4 z^4}{(z - a_2)^2(z - a_3)^2}$ | $z = a_1 + e^{\tfrac{x - x_0}{\sigma}}$, LB [14] | $_2F_1$ [8] |
| 10 | $\tfrac{1}{2}, \tfrac{1}{2}, \tfrac{1}{2}$ | $\dfrac{V_0 + V_1 z + V_2 z^2 + V_3 z^3 + V_4 z^4}{(z - a_1)(z - a_2)(z - a_3)}$ | $z = a_1 + (a_2 - a_1)/\mathrm{sn}\left(\dfrac{\sqrt{a_2 - a_1}(x_0 - x)}{2\sigma}\bigg|\dfrac{a_1 - a_3}{a_1 - a_2}\right)^2$, LB [14]<br>sn is the Jacobi elliptic sine function | -- |
| 11 | $\tfrac{1}{2}, \tfrac{1}{2}, 0$ | $\dfrac{V_0 + V_1 z + V_2 z^2 + V_3 z^3 + V_4 z^4}{(z - a_1)(z - a_2)(z - a_3)^2}$ | $z = a_1 + \dfrac{1}{4}\left(e^{\tfrac{x - x_0}{2\sigma}} + (a_2 - a_1)e^{-\tfrac{x - x_0}{2\sigma}}\right)^2$, LB [14] | $_2F_1$ [10] |

Table 1. Eleven independent general Heun potentials. Five potentials are given explicitly as $V(z(x))$, the others are in general defined parametrically as a pair of functions $x(z), V(z)$. "LB" refers to the paper by Lemieux and Bose. Potentials possessing classical exactly integrable hypergeometric sub-potentials are marked in the last column by $_2F_1$. The three potentials possessing conditionally integrable hypergeometric sub-potentials for which each of the fundamental solutions involves a single Gauss hypergeometric function are marked by "CIP".

Some of the potentials for the canonical specification of the singularities $a_{1,2,3} = 0,1,a$ have been previously presented by Lemieux and Bose [14]. Though they have listed eight potentials in Table 1 of [14], however, these are particular cases of only four independent general Heun potentials. Precisely, the first four rows of that table present the Heun potential $m_{1,2,3} = (1/2,1/2,0)$ subject to the specifications for $x_0,\sigma,a$ given as $(0,1/(2\alpha),c)$, $(i\pi/(2\alpha),1/\alpha,(1-ic)/2)$, $(0,1/(2\alpha),c/(c-1))$ and $(0,1/(2\alpha),1/(c-1))$, respectively. The fifth and sixth rows present $x_0 = i\pi\sigma$ and $x_0 = 0$ specifications of the case $m_{1,2,3} = (1,1,0)$ with $\sigma = 1/(2\alpha), a = (1+c)/2$, the seventh row is the $x_0,\sigma,a = -(i\pi + \ln g)\sigma, 1/\alpha, 1-c$ specification of the potential $m_{1,2,3} = (1,0,0)$, and the last row is the specification $x_0,\sigma,a = -\left(\ln\sqrt{-g}\right)\sigma, 2/\alpha, -1$ of the potential $m_{1,2,3} = (1,1,1)$. It should be said that in their table Lemieux and Bose limited themselves to potentials defined by simple and explicit functions. However, they note that "*removing these limitations would significantly lengthen the table. In particular, we would get a potential defined by elliptic functions. The Jacobian form of the Lamé equation would be a particular example*" [14]. Thus, the fifth explicit potential with $m_{1,2,3} = 1/2$ was also explicitly indicated by Lemieux and Bose.

## 4. The solution of the Schrödinger equation

For all potentials, not only the ones listed in Table 1, using the auxiliary parameters $r_{0,1,2,3,4}$ and $v_{0,1,2,3,4}$ calculated through the definitions (14),(15), the solution of the Schrödinger equation (1) is explicitly written as

$$\psi = (z-a)^{\alpha_1} z^{\alpha_2} (z-1)^{\alpha_3} H_G(a_1,a_2,a_3; q; \alpha,\beta,\gamma,\delta,\varepsilon; z), \qquad (16)$$

where the general Heun function's parameters $\alpha,\beta,\gamma,\delta,\varepsilon$ and $q$ are given by the equations

$$\gamma + \delta + \varepsilon = \alpha + \beta + 1, \qquad (17)$$

$$\gamma = 2\alpha_1 + m_1, \quad \delta = 2\alpha_2 + m_2, \quad \varepsilon = 2\alpha_3 + m_3, \qquad (18)$$

$$\alpha\beta = (\alpha_1+\alpha_2+\alpha_3)^2 - (\alpha_1+\alpha_2+\alpha_3)(m_1+m_2+m_3-1) + \frac{2m}{\hbar^2}(Er_4 - v_4), \qquad (19)$$

$$\begin{aligned}
q = &\, a_1\left((\alpha_2+\alpha_3)(\alpha_2+\alpha_3+m_2+m_3-1) - \alpha_1(\alpha_1+m_1-1)\right) + \\
&+ a_2\left((\alpha_1+\alpha_3)(\alpha_1+\alpha_3+m_1+m_3-1) - \alpha_2(\alpha_2+m_2-1)\right) + \\
&+ a_3\left((\alpha_1+\alpha_2)(\alpha_1+\alpha_2+m_1+m_2-1) - \alpha_3(\alpha_3+m_3-1)\right) + \\
&- \frac{2m}{\hbar^2}\left(E(r_3 + (a_1+a_2+a_3)r_4) - (v_3 + (a_1+a_2+a_3)v_4)\right)
\end{aligned} \qquad (20)$$

and the exponents $\alpha_{1,2,3}$ of the pre-factor are defined by the equations

$$(1-m_i-\alpha_i)\alpha_i \prod_{n=1,n\neq i}^{3}(a_i-a_n)^2 = \frac{2m}{\hbar^2}(Er(a_i)-v(a_i)), \quad i=1,2,3. \tag{21}$$

The general Heun function $H_G$ can be constructed as an expansion in terms of powers of $z$ [15-16] or in terms of familiar special functions of the hypergeometric class [34-41]. For the canonical specification of the singularities as $a_{1,2,3}=(0,1,a)$ [15-16] the Frobenius power series in the vicinity of the singular point $z=0$:

$$H(a,q;\alpha,\beta;\gamma,\delta;z) = z^{\mu} \sum_{n=0}^{+\infty} c_n z^n, \tag{22}$$

is governed by the following three-term recurrence relation for the successive coefficients $c_n, c_{n-1}, c_{n-2}$ of the expansion [15-16]

$$R_n c_n + Q_{n-1} c_{n-1} + P_{n-2} c_{n-2} = 0, \tag{23}$$

where
$$R_n = a(\mu+n)(\mu+n-1+\gamma), \tag{24}$$

$$Q_n = -q - (\mu+n)\big((\mu+n-1+\gamma+\delta+\varepsilon)(1+a) - a\varepsilon - \delta\big), \tag{25}$$

$$P_n = (\mu+n+\alpha)(\mu+n+\beta) \tag{26}$$

and $\mu=0$ or $\mu=1-\gamma$. The recursion (23) is evolved for $n\geq 1$ with the initial conditions $c_0 \neq 0$, $c_{-1}=c_{-2}=0$.

Many *quasi-exactly solvable* models [42] are constructed via the right-hand side termination of this series. Such a termination is achieved at some $n=N$, $N=1,2,3,...$ if one demands $c_{N+1}=c_{N+2}=0$. These conditions are satisfied if $P_N=0$, that is $\mu+\alpha=-N$ or $\mu+\beta=-N$, and additionally $Q_N c_N + P_{N-1} c_{N-1} = 0$. The last equation, which is a polynomial equation of the order $N+1$ for the accessory parameter $q$, is an equation that may serve as a spectral equation for energy quantization.

Note that the convergence radius of the series (22) is $\min\{|a|,1\}$ so that at $|a|<1$ it does not cover the whole interval between the singular points $z=0$ and $z=1$ that are often involved in physical problems. In such cases more advantageous may be the application of other expansion functions instead of simple powers. One may use expansions in terms of the functions of the hypergeometric class [34-39] or expansions in terms of higher transcendental functions such as the Goursat generalized hypergeometric function or the Appell hypergeometric function of two variables [40,41]. The convergence region for these series may be other than a circle [34]. Below we apply a particular expansion in terms of the Gauss hypergeometric functions [39] that leads to closed form solutions of the Schrödinger equation for which each of the fundamental solutions composing the general solution is written as a linear combination of a finite number of hypergeometric functions. This expansion reads:

$$u = \sum_{n=0}^{+\infty} c_n \cdot {}_2F_1(\alpha, \beta; \gamma_0 - n; z), \qquad (27)$$

where the coefficients $c_n$ obey the following three-term recurrence relation:

$$R_n c_n + Q_{n-1} c_{n-1} + P_{n-2} c_{n-2} = 0 \qquad (28)$$

with

$$R_n = \frac{a}{\gamma_0 - n}(\gamma - \gamma_0 + n)(\alpha - \gamma_0 + n)(\beta - \gamma_0 + n), \qquad (29)$$

$$Q_n = (1-a)(\varepsilon + \gamma - \gamma_0 + n)(\gamma_0 - n - 1) + a(\gamma - \gamma_0 + n)(\alpha + \beta - \gamma_0 + n) + \alpha\beta a - q, \qquad (30)$$

$$P_n = (a-1)(\varepsilon + \gamma - \gamma_0 + n)(\gamma_0 - n - 1) \qquad (31)$$

and $\gamma_0 = \gamma$ or $\alpha$ or $\beta$. The right-hand side termination of this series results in conditionally or quasi-exactly integrable sub-potentials. The termination happens at $n = N = 1, 2, 3...$ if

$$\varepsilon \text{ or } \varepsilon + \gamma - \alpha \text{ or } \varepsilon + \gamma - \beta = -N \qquad (32)$$

for $\gamma_0 = \gamma$ or $\alpha$ or $\beta$, respectively, and $c_{N+1} = 0$. The second condition for termination of the series, which is equivalent to the equation $Q_N c_N + P_{N-1} c_{N-1} = 0$, is a polynomial equation of the order $N+1$ for the accessory parameter $q$ having in general $N+1$ solutions. This equation imposes an additional restriction on the parameters involved in the potential.

## 5. Exactly and conditionally exactly solvable hypergeometric sub-potentials

Many particular cases of the presented general Heun potentials admitting solution in terms of the Gauss hypergeometric functions have been reported in the past (see, e.g., [43-60]). It is worth to separate the whole set into the cases for which the Heun function is written as a product of an elementary function and a *single* hypergeometric function and the more advanced cases for which the fundamental solutions are given through irreducible linear combinations of *two or more* hypergeometric functions.

Consider the sub-potentials for which the Heun function and thus the solution of the Schrödinger equation involves only one hypergeometric function. By shifting and scaling the independent variable as $z \to a_1 + (a_2 - a_1)z$ and setting $(a_3 - a_1)/(a_2 - a_1) = a$, the general Heun equation is always written in the *canonical* form with $a_{1,2,3} = 0, 1, a$ [15,16]. As it is then immediately seen from equation (10), the canonical general Heun equation is reduced to the Gauss hypergeometric equation if $\varepsilon = 0$ and $q = a\alpha\beta$. Examining the possibility of satisfying these two conditions through equations (17)-(21), one reveals that this is achieved in ten cases if $m_{1,2,3}$ obey the inequalities $-1 \leq m_{1,2} \leq 1 \cup -1 \leq m_3 \leq 0$ and $1 \leq m_1 + m_2 + m_3 \leq 3$. Because of the symmetry with respect to the transposition $m_1 \leftrightarrow m_2 \cup z \leftrightarrow 1-z$, the number of the independent cases is reduced to seven. Four of these independent cases (lines 3,7,9,11 of Table 1), the ones for which the inverse transformation $z(x)$ is always written in terms of elementary functions, present the classical five-parametric exactly integrable hypergeometric potentials, either the Eckart potential [8] or the Pöschl-Teller potential [10]. The other three cases (lines 4,5,8 of Table 1) have conditionally integrable hypergeometric sub-potentials. The latter cases are marked in the last column of Table 1 by "CIP". The corresponding restrictions imposed on the parameters of the potential for each of these three cases for the canonical choice $a_{1,2,3} = (0,1,a)$ are presented in Table 2.

We conclude this section by noting that the general Heun equation is reduced to the Gauss hypergeometric equation in several other ways. For instance, straightforward reductions are achieved through simple linear transformations of the argument if $(\delta, q) = (0, \alpha\beta)$ or if $(\gamma, q) = (0, 0)$ (for the canonical equation with $a_{1,2,3} = 0, 1, a$). Many reductions applying rational transformations of the independent variable are presented in [61-63]. The less restrictive cases among all such reductions are the ones which apply quadratic transformation of the argument. It is however checked that the latter reductions merely reproduce the known two classical Gauss hypergeometric potentials by Eckart [8] and Pöschl-Teller [10]. In Table 1 we indicate three (of many) such cases (lines 1,5 and 6). This is to show that many of the general Heun potentials present distinct generalizations of the classical hypergeometric potentials.

| N | $m_{1,2,3}$ | Restrictions |
|---|---|---|
| 4 | $1, 1, -\dfrac{1}{2}$ | $V_0 = \dfrac{(a-1)(7a-1)a^2\hbar^2}{32m\sigma^2} + a^2\left(V_2 + a(2V_3 + 3aV_4)\right)$, $V_1 = \dfrac{3(1-a)(2a-1)a\hbar^2}{16m\sigma^2} - a\left(2V_2 + a(3V_3 + 4aV_4)\right)$ |
| 5 | $1, 1, -1$ | $V_0 = \dfrac{(a-1)(5a-1)a^2\hbar^2}{8m\sigma^2} + a^2\left(V_2 + a(2V_3 + 3aV_4)\right)$, $V_1 = \dfrac{(1-a)(2a-1)a\hbar^2}{2m\sigma^2} - a\left(2V_2 + a(3V_3 + 4aV_4)\right)$ |
| 8 | $1, \dfrac{1}{2}, -\dfrac{1}{2}$ | $V_0 = \dfrac{(a-1)(9a-1)a^2\hbar^2}{32m\sigma^2} + a^2\left(V_2 + a(2V_3 + 3aV_4)\right)$, $V_1 = \dfrac{(1-a)(7a-3)a\hbar^2}{16m\sigma^2} - a\left(2V_2 + a(3V_3 + 4aV_4)\right)$ |

Table 2. The three conditionally integrable potentials (CIP) for which each of the fundamental solutions involves a single hypergeometric function. The number in the first column indicates the number of the independent Heun potential from Table 1 to which the particular hypergeometric sub-potential belongs.

## 6. Other Gauss hypergeometric sub-potentials

It is known that there exist several potentials for which the general solution of the Schrödinger equation involves fundamental solutions that are given through irreducible linear combinations of two or more hypergeometric functions (see, e.g., [24-27,31,32,37,43]). These potentials, which may be exactly [24-27] or conditionally exactly [31,32,37,43] integrable, can be conveniently derived by termination of the series solutions of the Heun equations in terms of functions of the hypergeometric class.

As it comes to the general Heun potentials discussed here, a representative example is the third exactly solvable hypergeometric potential [27]

$$V(x) = V_0 + \frac{V_1}{\sqrt{1 + e^{2(x-x_0)/\sigma}}} \tag{33}$$

for which each of the two fundamental solutions that compose the general solution of the Schrödinger equation presents an irreducible combination of two Gauss hypergeometric functions $_2F_1$. This potential belongs to the case $m_{1,2,3} = (1,1,-1)$ with specialization of the singularities of the general Heun equation as $a_{1,2,3} = (-1,1,0)$.

A conditionally exactly integrable example belonging to the same Heun potential and for which the fundamental solutions are again written through linear combinations of two Gauss hypergeometric functions is the potential recently reported by A. López-Ortega [43]:

$$V = \left( \frac{A^2 e^x}{e^x + 1} + \frac{A e^{x/2}}{(e^x + 1)^{3/2}} \right) / \frac{2m}{\hbar^2}. \tag{34}$$

We note that from the point of view of the super-symmetric quantum mechanics [64] the third exactly solvable hypergeometric potential (33) is the super-potential for the conditionally solvable partner potentials (34) constructed by putting $A \to \pm A$.

We here present a generalization of the potentials (33) and (34) written as

$$V = V_0 + \frac{V_1}{z} + \frac{2m\sigma^2 V_3^2/\hbar^2}{z^2} + \frac{V_3}{z^3}, \quad z = \sqrt{1 + e^{2(x-x_0)/\sigma}}, \tag{35}$$

where $V_{0,1,2}$ and $x_0, \sigma$ are arbitrary real or complex parameters. The third hypergeometric potential (33) is obtained by putting $V_3 = 0$ and the potential (34) by López-Ortega is reproduced by choosing $V_1 = -V_3 = A\hbar^2/(4m)$. The potential shapes are shown in Fig.2.

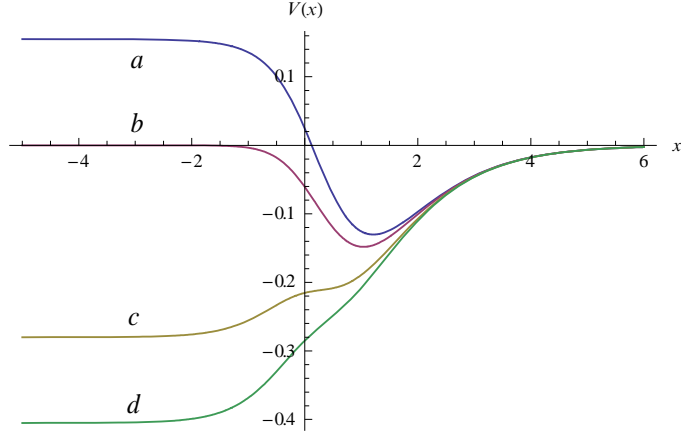

Fig. 2. Conditionally integrable potential (35). $V_3 = -1.05, -1, -0.9, -0.85$ - curves a,b,c,d, respectively, $V_0 = x_0 = 0$, $V_1 = -1$, $\hbar = m = \sigma = 1$.

It is readily checked by direct substitution that the Schrödinger equation (1) for this potential with arbitrary parameters $V_{0,1,2}$ and $x_0, \sigma$ admits a fundamental solution written as

$$\psi(x) = (z+1)^{\alpha_1} (z-1)^{\alpha_2} u(z), \qquad (36)$$

$$u(z) = {}_2F_1\left(\alpha-1, \beta; \gamma-1; \frac{z+1}{2}\right) + \frac{\alpha_2 - \alpha_1 + \beta z - 2m\sigma^2 V_3/\hbar^2}{2(\gamma-1)} {}_2F_1\left(\alpha, \beta+1; \gamma; \frac{z+1}{2}\right), \qquad (37)$$

where the involved parameters are given as

$$(\alpha, \beta, \gamma) = (\alpha_1 + \alpha_2 - \alpha_0, \alpha_1 + \alpha_2 + \alpha_0, 1 + 2\alpha_1), \qquad (38)$$

$$\alpha_0 = \pm\sqrt{\frac{-2m\sigma^2}{\hbar^2}(E - V_0)},$$

$$\alpha_{1,2} = \left(\pm\sqrt{\frac{-m\sigma^2}{2\hbar^2}\left(E - V_0 + V_1 + V_3 - \frac{2m\sigma^2 V_3^2}{\hbar^2}\right)}, \pm\sqrt{\frac{-m\sigma^2}{2\hbar^2}\left(E - V_0 - V_1 - V_3 - \frac{2m\sigma^2 V_3^2}{\hbar^2}\right)}\right). \quad (39)$$

Here any combination for the signs of $\alpha_{0,1,2}$ is applicable. We note that by choosing different combinations of the signs one constructs different fundamental solutions and thus, by taking any pair of independent ones, arrives at the general solution of the Schrödinger equation.

This solution is derived by termination of the above series (27)-(31) for the parameters given by equations (17)-(21) by applying the termination conditions with $\varepsilon = -1$ (see equation (32)). The technical lines are exactly the same as those described in [27] except that in the very last step one allows the auxiliary parameters $v_{0,1,2,3,4}$ to be dependent.

With the potential that we have just presented as well as the three ones from Table 2, the reported set of conditionally integrable hypergeometric sub-potentials as well as some quasi-exactly solvable cases of the general Heun potentials are summarized in Table 3. Note that the potential (35) and the potential (34) by López-Ortega are here listed as members of the case $m_{1,2,3}=(1,1,1)$ with the singularities of the Heun equation taken as $a_{1,2,3}=(0,1,-1)$ because they have such an alternative representation. This is to demonstrate that the same Gauss hypergeometric sub-potential may have different general Heun representations.

We would like to conclude by noting that the list of conditionally integrable hypergeometric potentials by no means is restricted to Table 3. Many other such potentials (presumably, infinitely many) are derived by checking the termination conditions for the series expansions of the general Heun functions in terms of the ordinary or generalized hypergeometric functions.

| N | $m_{1,2,3}$ | Potential and coordinate transformation | $a_{1,2,3}$ and $x_0,\sigma$ | Note |
|---|---|---|---|---|
| 1 | 1, 1, 1 | $V = \dfrac{W_1 e^x}{e^x+1} + \dfrac{W_2 e^{x/2}}{(e^x+1)^{3/2}}$, $z = \dfrac{1}{\sqrt{1+e^{-x}}}$ | $a_{1,2,3}=0,1,-1$ $x_0,\sigma = i\pi,-2$ | A. López-Ortega Eq.(4) [43] This work CIP |
| 2 | $1,1,\dfrac{1}{2}$ | $V(z) = W_1 y^2 + W_2 y^4 + W_3 y^6$, $y \to \sqrt{z}$ $x(z) = \dfrac{1}{\lambda^2}\left(\tanh^{-1} y - \sqrt{1-\lambda^2}\tanh^{-1}\left(\sqrt{1-\lambda^2}\, y\right)\right)$, $y \to \sqrt{z}$ | $a_{1,2,3}=\dfrac{1}{1-\lambda^2},1,0$ $x_0,\sigma = 0, \dfrac{1/2}{1-\lambda^2}$ | J.N. Ginocchio Eq.(2.7) [44] CIP |
| 3 | 1, 1, 0 | $V = \dfrac{W_1}{\cosh^2(\alpha x)} + \dfrac{W_2}{\cosh^2(\alpha(x+a))} + W_3 \tanh(\alpha x)$, $z = \dfrac{1}{1+e^{2\alpha x}}$ | $a_{1,2,3}=0,1,\dfrac{1+\coth(\alpha a)}{2}$ $x_0,\sigma = i\pi\sigma, 1/(2\alpha)$ | B. Stec Eq.(1.2) [45] CIP |
| 4 | $1,1,-\dfrac{1}{2}$ | $V = \dfrac{\hbar^2}{2m}\dfrac{(3/4)z(1-z)+h_0(1-z)+h_1 z^{1/2}}{b_0 z + c_0} + \dfrac{\hbar^2}{28m}\left(3\left(\dfrac{z''(x)}{z'(x)}\right)^2 - 2\dfrac{z'''(x)}{z'(x)}\right)$, $z'(x) = \dfrac{2z(1-z)}{\sqrt{b_0 z + c_0}}$ | $a_{1,2,3}=0,1,-c_0/b_0$ $x_0,\sigma = 0,\sqrt{7b_0/4}$ Table 2 | C. Grosche Eqs.(31),(48) [46] $h_1=0$ This work CIP |
| 5 | 1, 1, −1 | $V = \dfrac{W_1\left((1-\sigma)+(1+\sigma)e^{-x}\right)}{\sqrt{1+e^{2x/\sigma}}} + \dfrac{W_2}{1+e^{2x/\sigma}} + \dfrac{-3\hbar^2/(8m\sigma^2)}{(1+e^{2x/\sigma})^2}$, $z = \sqrt{1+e^{2x/\sigma}}$ | $a_{1,2,3}=-1,1,0$ $x_0,\sigma = 0,\mp 1$ Table 2 | C. Grosche Eqs.(37),(51) [47] This work CIP |
| 6 | $1,\dfrac{1}{2},\dfrac{1}{2}$ | | | $_2F_1(x^2)$ [10] |
| 7 | $1,\dfrac{1}{2},0$ | $V = \operatorname{sech}^2(\alpha x)\left(W_1 + \dfrac{W_2}{a - \tanh^2(\alpha x)} + \dfrac{W_3}{\left(a-\tanh^2(\alpha x)\right)^2}\right)$, $z = \operatorname{sech}^2(\alpha x)$ | $a_{1,2,3}=0,1,1-a$ $x_0,\sigma = 0, i/(2\alpha)$ | L. Trlifaj Eq.(18) [49] CIP |

| | | | | | |
|---|---|---|---|---|---|
| 8 | $1, \frac{1}{2}, -\frac{1}{2}$ | $V(z) = \dfrac{W_1 + W_2 y^2 + W_3 y^4}{(a^2 + y^2)^3} + \dfrac{cy}{a^2 + y^2}$, $y \to \sqrt{z}$ <br><br> $x(z) = \sinh^{-1}\left(\dfrac{\sqrt{z}}{a}\right) - \sqrt{1-a^2}\, \tanh^{-1}\left(\dfrac{\sqrt{1-a^2}\,\sqrt{z}}{\sqrt{z+a^2}}\right)$ | $a_{1,2,3} = -1, 0, -a^2$ <br> $x_0 = \left(\sqrt{1-a^2} - 1\right)\ln a$ <br> $\sigma = 1/2$ | R. Milson [50] Eq.(23) $c = 0$ CIP |
| | | $V(z) = \dfrac{W_1 + W_2 y + W_3 y^2}{(\delta + 1 - y^2)^3} + \dfrac{2C\Lambda y}{\delta + 1 - y^2}$, $y \to \sqrt{z}$ <br><br> $x(z) = \dfrac{1}{\sqrt{C}}\left(\tan^{-1}\left(\dfrac{\sqrt{z}}{\sqrt{\delta+1-z}}\right) + \sqrt{\delta}\,\tanh^{-1}\left(\dfrac{\sqrt{\delta}\sqrt{z}}{\sqrt{\delta+1-z}}\right)\right)$ | $a_{1,2,3} = 1, 0, \delta + 1$ <br> $x_0 = -\dfrac{i + \sqrt{\delta}}{2\sqrt{C}}\ln(-1-\delta)$ <br> $\sigma = i/(2\sqrt{C})$ | G. Lévai [51] Eq.(17) $\Lambda = 0$, CIP |
| | | $V(z) = \dfrac{W_0 + W_1 z + V_2 z^2 + V_3 z^3 + V_4 z^4}{(z-1)(z-a)^3}$ <br><br> $\dfrac{x - x_0}{\sigma} = \ln\left(2\sqrt{(z-1)(z-a)} - a + 2z - 1\right)$ <br> $- \sqrt{a}\left(\ln(-z) - \ln\left(2\sqrt{a}\sqrt{(z-1)(z-a)} - (a+1)z + 2a\right)\right)$ | $a_{1,2,3} = 0, 1, a$ <br> Table 2 | C. Grosche Eqs.(55),(70) [46] $h_1 = 0$ This work CIP |
| 9 | 1, 0, 0 | | | ${}_2F_1$ [8] |
| 10 | $\frac{1}{2}, \frac{1}{2}, \frac{1}{2}$ | $V = \dfrac{V_0 + V_1 z + V_2 z^2 + V_3 z^3 + V_4 z^4}{z(z-1)(z-a)}$, $z = 1/\mathrm{sn}\left(\dfrac{x_0 - x}{2\sigma}\Big| a\right)^2$ | $a_{1,2,3} = 0, 1, a$ <br> $x_0, \sigma$ - arbitrary | Darboux-Treibich-Verdier [52-54], QES, CIP |
| 11 | $\frac{1}{2}, \frac{1}{2}, 0$ | $V = \dfrac{W_1}{c + \sin(\alpha x)} + \dfrac{W_2}{(c + \sin(\alpha x))^2} + W_3 \sec^2(\alpha x) + W_4 \sec(\alpha x)\tan(\alpha x)$ <br> $z = \cos^2(\alpha x/2 - \pi/4)$ | $a_{1,2,3} = 0, 1, (1-c)/2$ <br> $x_0, \sigma = \pi/(2\alpha), i/\alpha$ | C. Quesne Eq.(11) [48] CIP |
| | | $V = \dfrac{W_1}{1 + c\cos^2(\alpha x)} + \dfrac{W_2}{(1 + c\cos^2(\alpha x))^2}$, $z = \cos^2(\alpha x)$ | $a_{1,2,3} = 0, 1, -1/c$ <br> $x_0, \sigma = 0, i/(2\alpha)$ | Eq.(30) [55] QES, CIP |
| | | $V = \dfrac{W_1}{1 + c\cosh^2(\alpha x)} + \dfrac{W_2}{(1 + c\cosh^2(\alpha x))^2} + \dfrac{W_3}{\cosh^2(\alpha x)}$, $z = \cosh^2(\alpha x)$ | $a_{1,2,3} = 0, 1, -1/c$ <br> $x_0, \sigma = 0, 1/(2\alpha)$ | Eq.(1) [56] Eq.(33) [57] Eq.(62) [58] QES, CIP |

Table 3. Several appearances of the general Heun potentials. The particular values of (arbitrary or fixed) constants $W_{1,2,3}$ adopted in the corresponding references are readily determined by direct comparison. "CIP" denotes the cases having sub-potentials that are conditionally integrable in terms of the Gauss hypergeometric functions, "QES" denotes quasi-exactly solvability.

## 7. Discussion

The general Heun equation is an equation with remarkably wide covering. It suggests a very large variety of possible analytically solvable cases for different physical problems that are reduced to the second-order linear differential equations.

However, the general Heun function is a rather complicated mathematical object. The reductions of this function to its immediate predecessor, that is to the familiar Gauss hypergeometric function $_2F_1$, have been discussed by many authors (see, e.g., [34-39,61-63]). An observation here is that the direct one-term Heun-to-hypergeometric reductions [61-63] suffer the essential disadvantage that for such reductions one should impose rather severe restrictions on the involved parameters, namely, three or more conditions. It turns out that these conditions are such that they are either not satisfied by the Heun potentials or, if satisfied, produce very restrictive potentials. The less restrictive reduction is achieved for the specification $\varepsilon = \delta$, $a = -1$ and $q = 0$ (or for two equivalent modifications of this specification achieved if $\gamma = \delta$ and $a = 1/2$, $q = \alpha\beta/2$ or if $\gamma = \varepsilon$ and $a = 2$, $q = \alpha\beta$ [61-63]), when three parameters out of total six of the canonical Heun equation are already specified so that only three parameters are left free. It is further checked that these specifications merely produce the known hypergeometric potentials, either the one by Eckart [8] or that by Pöschl and Teller [10].

Considerably more advanced are the terminations of the series expansions of the general as well as confluent Heun functions in terms of the functions of the hypergeometric class [34-39]. For comparison, in these cases only two restrictions are imposed on the involved parameters and, notably, these restrictions are such that in many cases they are satisfied. In this way the exact solutions for the inverse square root [24], the Lambert-W step [25] and the Lambert-W singular [26], as well as for the third independent Gauss hypergeometric [27] potentials have been recently derived. We have presented here another example of a conditionally exactly integrable potential for which the solution is written through fundamental solutions each of which is an irreducible combination of two Gauss hypergeometric functions. This is a conditionally integrable generalization of the third exactly solvable hypergeometric potential [27].

There have been many studies discussing the reducibility of the Schrödinger equation to the Heun class of equations (see, e.g. [15-17,23-27,33,37,65] and references therein). Discussing the case of the general Heun equation, we have shown that there exist in total 35 separate choices for the coordinate transformation, each leading to a ten-parametric potential (eight-parametric if the canonical form of the general Heun equation is considered). We note that analogous discretization of solvable models is observed for the Schrödinger equation [23], as well as for other wave equations including the Dirac and Klein-Gordon equations [66] and the quantum two-state problem [28-30] also in the confluent Heun cases.

Because of the symmetry of the general Heun equation with respect to the transposition of its singularities, only 11 potentials are independent in the sense that these potentials cannot be derived from each other by specifications of the involved parameters. Seven of the independent potentials present distinct generalizations of the classical ordinary hypergeometric potentials. Two of the remaining four potentials possess conditionally integrable hypergeometric sub-cases such that the solution is written through a single Gauss hypergeometric function, while two others have quasi-exactly solvable sub-potentials. There exist many sub-potentials for which the solution is written as a linear combination of a finite number of hypergeometric functions. We conclude by noting once more that the Heun potentials cover a considerably large set of physical problems as compared with the hypergeometric ones. Some of the non-hypergeometric representatives of the Heun potentials have been discussed by many authors on several occasions starting from the seminal works by Morse and E.C.G. Stückelberg [11] and Manning [12].

**Acknowledgments**

This research has been conducted within the scope of the International Associated Laboratory IRMAS (CNRS-France & SCS-Armenia). The work has been supported by the Armenian State Committee of Science (SCS Grant No. 15T-1C323), Armenian National Science and Education Fund (ANSEF Grant No. PS-4558) and the project "Leading Russian Research Universities" (Grant No. FTI_24_2016 of the Tomsk Polytechnic University).